\documentclass[preprint,superscriptaddress,amsmath,amssymb]{revtex4}
\usepackage{graphicx}
\usepackage{bm}
\usepackage{amsmath,amssymb}

\begin{document}

\title{Blind topological measurement-based quantum computation} 

\author{Tomoyuki Morimae*}
\affiliation{
Controlled Quantum Dynamics Theory Group, Imperial College London, London SW7 2AZ, United Kingdom
}
\affiliation{
Laboratoire d'Analyse et de Math\'ematiques Appliqu\'ees, 
Universit\'e Paris-Est Marne-la-Vall\'ee, 77454 Marne-la-Vall\'ee
Cedex 2, France
}

\author{Keisuke Fujii}
\affiliation{Graduate School of Engineering Science, Osaka University,
Toyonaka, Osaka 560-8531, Japan}
\date{\today}
            
\maketitle  

{\bf
Blind quantum computation 
is a novel secure quantum computing
protocol which enables Alice,
who does not have sufficient quantum technology
at her disposal,
to delegate her quantum computation
to Bob, who has a fully-fledged quantum computer
in such a way that Bob cannot learn anything
about Alice's input, output, and algorithm.
A recent proof-of-principle experiment demonstrating blind quantum computation
in an optical system
has raised new challenges regarding the scalability  
of blind quantum computation
in realistic noisy conditions.
Here we show that  
fault-tolerant blind quantum computation
is possible in a 
topologically-protected manner using the Raussendorf-Harrington-Goyal
scheme.
The error threshold of our scheme is
$4.3\times10^{-3}$, 
which is comparable that ($7.5\times10^{-3}$) 
of non-blind topological quantum
computation.
Since the error per gate of the order $10^{-3}$
was already achieved in some experimental systems, 
our result implies that secure cloud quantum computation
is within reach.
}


In classical computing, the problem of ensuring the communication
between a server and a client in secure is highly
non-trivial.
For example, 
Abadi, Feigenbaum and Killian 
showed 
that no NP-hard function
can be computed with encrypted data 
if unconditional security is required, 
unless the polynomial hierarchy collapses 
at the third level~\cite{Abadi}. 
Even restricting the security condition to be only computational,
the question of the possibility of 
the fully-homomorphic encryption has been a long-standing open question
for 30 years \cite{Gentry},
and still no practical method has been found.
Unconditionally secure fully-homomorphic encryption is still
an open problem.

On the other hand, 
in the quantum world, the situation is drastically different.
Broadbent, Fitzsimons and Kashefi proposed a  
quantum protocol~\cite{blindcluster},
so-called blind quantum computation~\cite{
blindcluster,Barz,TVE,Vedran,blind_bell,FK12}, which uses 
measurement-based quantum computation (MBQC)~\cite{cluster}.
In their protocol, Alice
has a classical computer and a quantum device that emits
randomly-rotated qubits. 
She does not have any 
quantum memory.
On the other hand, Bob has a fully-fledged quantum technology.
Alice and Bob share a classical channel and a quantum channel.
Their protocol runs as follows. First, Alice sends Bob many 
randomly-rotated qubits, and
Bob creates a graph state 
by applying C$Z$ gates among these qubits.
Second, Alice instructs Bob how to measure a qubit of the graph state.
Third, Bob measures the qubit according to Alice's instruction, and 
he returns the measurement result
to Alice. 
They repeat this classical two-way communication (i.e., 
the second and third steps) 
until the computation is finished.
It was shown~\cite{blindcluster}
that if Bob is honest, Alice can obtain the correct answer of
her desired quantum computation ({\it correctness}),
and that whatever evil Bob does, he cannot learn anything about
Alice's input, output, and algorithm ({\it blindness}).
Recently, this protocol has been experimentally demonstrated
in an optical system~\cite{Barz}.

Secure delegated computation is already in the practical phase
for classical computing, including smart phones, encrypted data 
retrieval~\cite{Song}, and
wireless sensor networks~\cite{wireless}.
When scalable quantum computers are realized, the need for delegated 
secure computation must be emphasized, 
since home-based quantum computers are arguably much more difficult 
to build than their classical counterparts. 
In order to implement blind quantum computation in a scalable manner, 
it is crucial to protect 
quantum computation from environmental noise.
In this paper, we show that a fault-tolerant 
blind quantum computation is possible in a
topologically protected manner.
We also calculate the error threshold,
$4.3\times10^{-3}$, for
erroneous preparation of the initial states, 
erroneous C$Z$ gates, and erroneous local measurements.
This is the first time that a concrete fault-tolerant scheme is
proposed for blind quantum computation, 
and that the error threshold is explicitly calculated. 
Furthermore, this threshold is not so different from
that, $7.5\times10^{-3}$, of non-blind
topological quantum computation~\cite{Raussendorf_PRL,Raussendorf_NJP}.
In other words, 
blind quantum computation is possible
with almost the same error threshold as
that of the non-blind version.
Since the error threshold of the order $10^{-3}$
was already achieved in some experimental systems~\cite{ion}, 
our result means that secure cloud quantum computation
is within our reach.
We further show that our protocol is also fault-tolerant against
the detectable qubit loss, such as a photon loss and an escape from
the qubit energy level.

\section{Results}
\subsection{Topologically-protected MBQC}
The topologically-protected measurement-based quantum computation
(TMBQC)~\cite{Raussendorf_PRL,Raussendorf_NJP,topo_experiment} 
is one of the most promising models of quantum computation.
In this model,
we first
prepare
the graph state
on the three-dimensional cubic lattice $\mathcal{L}$
whose elementary cell is given in Fig.~\ref{bcc3} (a).
We call this lattice 
${\mathcal L}$
the Raussendorf-Harrington-Goyal (RHG) lattice.
We next measure each qubit of this lattice in
$X$, $Z$, $T\equiv(X+Y)/\sqrt{2}$, or $Y$ basis.
These four types of measurements are sufficient for
the universal TMBQC
as is shown in Refs.~\cite{Raussendorf_PRL,Raussendorf_NJP}.
(More precisely, measurements in $X$ basis and $Z$ basis
can simulate
the topological braidings of defects in
the surface code~\cite{Kitaev}
which can implement the fault-tolerant Clifford gates,
and measurements in $T$ basis
and $Y$ basis 
can simulate
the preparations of magic states~\cite{magic}
which can implement the non-Clifford gates.
These magic states
are distilled~\cite{magic} by topologically protected
fault-tolerant Clifford gates which
are simulated by $X$ and $Z$ basis measurements.)
A small-size TMBQC has recently been experimentally demonstrated
in an optical system~\cite{topo_experiment}.

\subsection{Blind TMBQC}
Can we use TMBQC for the blind quantum computation?
Obviously, if Bob knows in which basis ($X$, $Z$, $T$,
or $Y$) he is doing the measurement on each qubit,
he can know Alice's algorithm.
However,
if Alice can have Bob do measurements
in such a way that
Bob cannot know in which basis ($X$, $Z$, $T$, or
$Y$) he is doing the measurement on each qubit, 
he cannot know Alice's algorithm.
How can Alice do that?
In fact, such a blind quantum computation
is possible if we consider the three-dimensional
lattice ${\mathcal L}'$ whose elementary cell
is given in Fig.~\ref{bcc3} (b)
where two extra qubits 
are added to each qubit of Fig.~\ref{bcc3} (a).
We call this lattice 
${\mathcal L}'$
the decorated RHG lattice.
The intuitive explanation of our idea is as follows:
First, it was shown~\cite{blindcluster}
that Alice can have Bob do the measurement
in $\{|0\rangle\pm e^{i\phi}|1\rangle\}$ basis for any 
$\phi\in\{\frac{k\pi}{4}~|~k=0,1,...,7\}$
on any qubit of a graph state which Bob has
in such a way that Bob cannot learn anything about $\phi$.
Second, it is easy to confirm that
a single-qubit measurement in
$X$, $Y$, $T$, or $Z$ basis
can be simulated
on the linear three-qubit cluster state
with only
$\{|0\rangle\pm e^{i\phi}|1\rangle\}$ basis measurements
(Fig.~\ref{aho}).
By combining these two facts, we notice that
if we decorate the RHG lattice as is shown in Fig.~\ref{bcc3} (b),
Bob can simulate
the measurement
in $X$, $Y$, $T$, and $Z$  basis only with
$\{|0\rangle\pm e^{i\phi}|1\rangle\}$ basis measurements, 
and he cannot know
which type of measurements ($X$, $Y$, $T$, or $Z$) 
he is simulating.

More precisely, our protocol runs as follows (Fig.~\ref{morimori}):

Step 1. Alice sends $N$
randomly-rotated single-qubit states $\{|\theta_j\rangle\}_{j=1}^N$
to Bob through the quantum channel, where
\begin{eqnarray*}
|\theta_j\rangle\equiv|0\rangle+e^{i\theta_j}|1\rangle,
\end{eqnarray*}
and
$\theta_j\in\{\frac{k\pi}{4}|k=0,1,...,7\}$ 
$(j=1,2,...,N)$
are random numbers.
$N$ is the total number of qubits used in the decorated RHG lattice
${\mathcal L}'$.
Alice remembers all random numbers $\Theta\equiv\{\theta_j\}_{j=1}^N$, 
and they are kept secret to Bob.

Step 2. Now Bob has $\{|\theta_j\rangle\}_{j=1}^N$. 
He places $|\theta_j\rangle$ on the $j$th vertex of the
decorated RHG lattice ${\mathcal L}'$ for all $j$ ($j=1,...,N$).
He then applies the C$Z$ gate on each red edge of the decorated
RHG lattice ${\mathcal L}'$.
Let us denote thus created $N$-qubit state
by $|\mathcal {C}_\Theta\rangle$.
Since 
\begin{eqnarray*}
|{\mathcal C}_\Theta\rangle
&=&
\Big(\bigotimes_{k,l} CZ_{k,l}\Big)
\Big(\bigotimes_{j=1}^Ne^{-iZ\theta_j/2}\Big)
|+\rangle^{\otimes N}\\
&=&\Big(\bigotimes_{j=1}^Ne^{-iZ\theta_j/2}\Big)
\Big(\bigotimes_{k,l} CZ_{k,l}\Big)
|+\rangle^{\otimes N},
\end{eqnarray*}
$|{\mathcal C}_\Theta\rangle$
is nothing but a rotated graph state on the
decorated RHG lattice ${\mathcal L}'$.
Here $CZ_{k,l}$ is the C$Z$ gate between $k$th and $l$th qubits.

Step 3. If Alice wants Bob to measure $j$th qubit of 
$|{\mathcal C}_\Theta\rangle$ 
in $\{|0\rangle\pm e^{i\phi_j}|1\rangle\}$ basis,
she calculates 
\begin{eqnarray*}
\delta_j\equiv\phi'_j+\theta_j+r_j\pi~~(\mbox{mod}~2\pi)
\end{eqnarray*}
on her classical computer.
Here, $r_j\in\{0,1\}$ is a random number,  
$\phi'_j\equiv(-1)^{s_j^X}\phi_j+\pi^{s_j^Z}$ (mod $2\pi$),
and $s_j^X,s_j^Z\in\{0,1\}$ are determined by the previous measurement
results
(this is the usual feed-forwarding in the one-way 
model~\cite{cluster}). 
Then Alice sends $\delta_j$ 
to Bob through the classical channel.

Step 4. Bob measures $j$th qubit in the 
$\{|0\rangle\pm e^{i\delta_j}|1\rangle\}$ basis, and
returns the result of the measurement to Alice through the
classical channel. 

Step 5. They repeat steps 3 and 4 with increasing $j$ until
they finish
the computation.
Note that Alice does the classical processing for
the error correction by using Bob's measurement results.

\subsection{Correctness}
\begin{figure}[htbp]
\begin{center}
\includegraphics[width=0.6\textwidth]{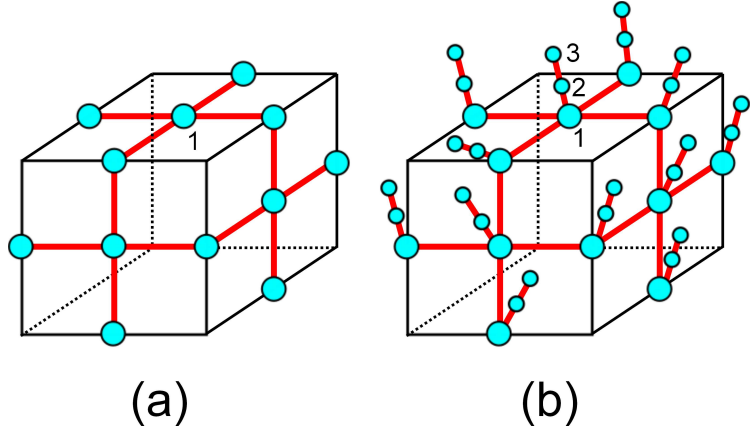}
\end{center}
\caption{
{\bf Elementary cells.}
(a): The elementary cell of the RHG lattice $\mathcal{L}$.
(b): The elementary cell of the decorated RHG lattice ${\mathcal L}'$. 
} 
\label{bcc3}
\end{figure}

\begin{figure}[htbp]
\begin{center}
\includegraphics[width=0.5\textwidth]{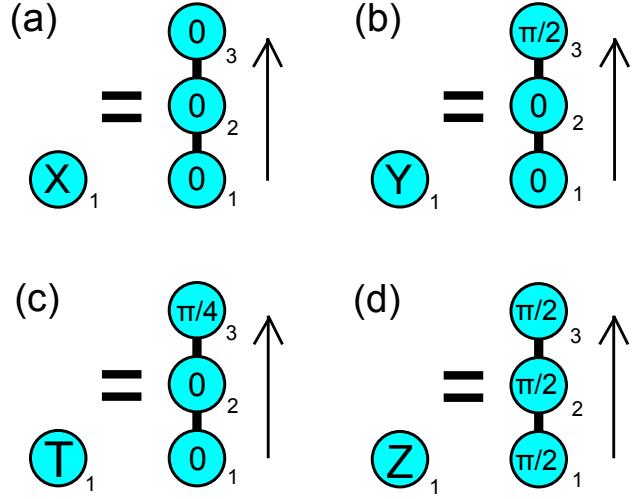}
\end{center}
\caption{
{\bf How the decorated lattice works.}
If we prepare the three-qubit cluster state
and measure each qubit 
in the numerical order 
in the 
$\{|0\rangle\pm e^{i\phi_j}|1\rangle\}$ basis
with 
$(\phi_1,\phi_2,\phi_3)=(0,0,0)$,
$(0,0,\pi/2)$, 
$(0,0,\pi/4)$, 
and
$(\pi/2,\pi/2,\pi/2)$,
we can simulate
single-qubit measurements in $X$, $Y$, $T$, and $Z$ basis,
respectively.
Each corresponds to (a), (b), (c), and (d).
} 
\label{aho}
\end{figure}

\begin{figure}[htbp]
\begin{center}
\includegraphics[width=0.8\textwidth]{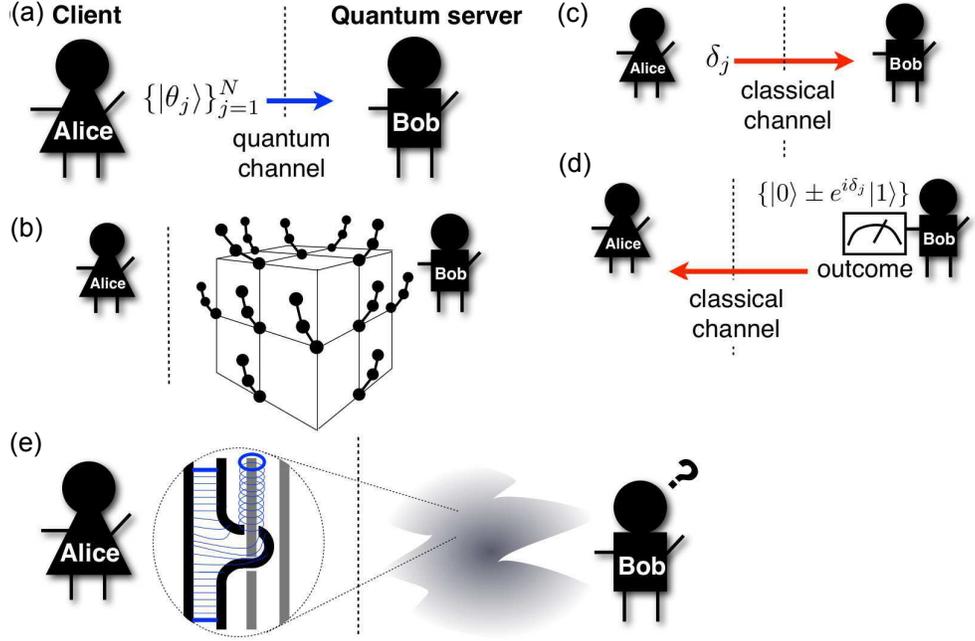}
\end{center}
\caption{
{\bf Topological blind protocol.}
(a) Alice sends Bob randomly rotated qubits.
(b) Bob creates the decorated RHG lattice.
(c) Alice sends Bob a classical message.
(d) Bob does the measurement and returns the result to Alice.
(e) Alice can hide her topological quantum computation from Bob.
} 
\label{morimori}
\end{figure}

Let us show that this protocol is correct.
In other words,
Alice and Bob can simulate the original TMBQC~\cite{Raussendorf_PRL,Raussendorf_NJP} 
on the decorated RHG lattice ${\mathcal L}'$ with only 
$\{|0\rangle\pm e^{i\phi}|1\rangle\}$ basis measurements
if Bob is honest.
Let us consider three qubits labeled with 1, 2, and 3,
in Fig.~\ref{bcc3} (b).
Let us assume that Bob
measures these three qubits in the numerical order
(i.e., from the bottom one to the top one)
in the
\begin{eqnarray*}
\{|0\rangle\pm e^{i\delta_j}|1\rangle\}
=
\Big\{e^{-i\theta_j Z/2}Z^{s^Z_j}X^{s^X_j}
\big(|0\rangle\pm e^{i\phi_j}|1\rangle\big)\Big\}
\end{eqnarray*}
basis ($j=1,2,3$)
with 
$(\phi_1,\phi_2,\phi_3)=(0,0,0)$.
By a straightforward calculation,
it is easy to show that 
such a sequence of measurements on the three qubits
is equivalent to the measurement
in $X$ basis on the qubit labeled with 1 in Fig.~\ref{bcc3} (a)
(also Fig.~\ref{aho} (a)).
(Note that $\theta_j$ in Bob's measurement basis is canceled since
$j$th qubit is pre-rotated by $\theta_j$ by Alice.
Furthermore, $r_j\pi$ causes just the flip of the measurement result.
Therefore, Bob effectively does 
$\{|0\rangle\pm e^{i\phi_j}|1\rangle\}$
basis measurement although he is doing 
$\{|0\rangle\pm e^{i\delta_j}|1\rangle\}$
basis measurement.)
In other words, our lattice ${\mathcal L}'$ can 
simulate $X$ basis measurements
on ${\mathcal L}$.
In a similar way,
if Bob does measurements
on the three qubits labeled with 1, 2, and 3 in Fig.~\ref{bcc3} (b)
in other angles, 
$(\phi_1,\phi_2,\phi_3)=(0,0,\pi/2)$,
$(0,0,\pi/4)$,
and
$(\pi/2,\pi/2,\pi/2)$,
they are equivalent to the $Y$, $T$, and $Z$
basis measurements on the qubit labeled with 1
in Fig.~\ref{bcc3} (a), respectively (Fig.~\ref{aho} (b), (c), and (d)).
In this way, our lattice ${\mathcal L}'$ can simulate $X$, $Z$, $T$,
and $Y$ basis measurements on ${\mathcal L}$.
In short, we have shown that 
$|{\mathcal C}_\Theta\rangle$ 
on the decorated RHG lattice ${\mathcal L}'$ 
can simulate the original
TMBQC~\cite{Raussendorf_PRL,Raussendorf_NJP} on 
the RHG lattice ${\mathcal L}$
solely with $\{|0\rangle\pm e^{i\phi}|1\rangle\}$ basis measurements.

\subsection{Blindness}
How about the blindness?
In our protocol, what Alice sends to Bob are
randomly rotated single-qubit states $\{|\theta_j\rangle\}_{j=1}^N$
and measurement angles 
$\{\delta_j\}_{j=1}^N$.
Without loss of generality, we can assume that the preparation of the input 
is included in the algorithm. Therefore, what Alice wants to hide are
the algorithm and the output.
We can show that
the conditional probability distribution of  
Alice's computational angles, 
given all the classical information Bob can obtain during the protocol, 
and given the measurement results of any POVMs which Bob may perform
on his system at any stage of the protocol,
is equal to the a priori probability distribution of
Alice's computational angles.
We can also show that
the final classical output is one-time padded
to Bob. (For a proof, see Methods.)

\subsection{Threshold}
\begin{figure}[htbp]
\begin{center}
\includegraphics[width=0.6\textwidth]{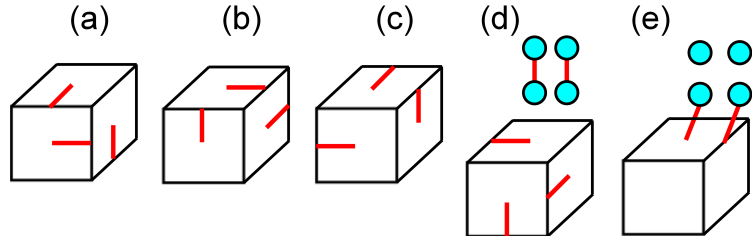}
\end{center}
\caption{
{\bf The stepwise creation of the resource state}.
$|\mathcal{C}_\Theta\rangle$
is created
from (a) to (e).
} 
\label{step}
\end{figure}

\begin{figure}[htbp]
\begin{center}
\includegraphics[width=0.6\textwidth]{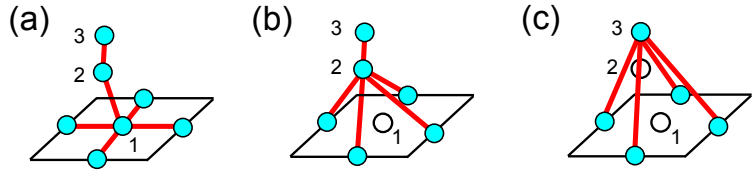}
\end{center}
\caption{
{\bf The measuring pattern.}
Three qubits are measured in the numerical order
from (a) to (c).
} 
\label{tower3}
\end{figure}

Finally, let us
calculate the fault-tolerant threshold.
As in Refs.~\cite{Raussendorf_PRL,Raussendorf_NJP},
we assume that errors occur during the preparation
of initial states 
$\{|\theta_j\rangle\}_{j=1}^N$, 
the applications of C$Z$ gates, and the local measurements.
The erroneous preparation of an initial state
is modeled by the perfect preparation followed by the 
partially
depolarizing noise 
with the probability $p_P$:
$(1-p_P)[I]+\frac{p_P}{3}([X]+[Y]+[Z])$,
where $[\bullet]$ indicates the super operator.
The erroneous local measurement is modeled by the perfect local measurement
preceded by the partially depolarizing noise with the probability $p_M$.
The erroneous C$Z$ gate is modeled by the perfect C$Z$ gate
followed by the two-qubit partially depolarizing noise 
with the probability $p_2$:
$(1-p_2)[I\otimes I]+
\frac{p_2}{15}([I\otimes X]+...+[Z\otimes Z])$.
Because of the rotational symmetry of the depolarizing noise,
we can replace $[A]$ 
with $[e^{-i\theta_j Z/2}A e^{i\theta_j Z/2}]$
when it acts on the $j$th qubit, 
and
$[A\otimes B]$  
with 
$[(e^{-i\theta_j Z/2}A e^{i\theta_j Z/2})\otimes
(e^{-i\theta_k Z/2}B e^{i\theta_k Z/2})]$
when it acts on the $j$th and $k$th qubits.
Here, $A,B=I,X,Y$ or, $Z$.
These replacements just correspond to the rotation of the local reference
frame of each qubit.
Then, if 
the measurement basis on the $j$th qubit ($j=1,...,N$) is
rotated by $e^{-i\theta_j Z/2}$,
the factor $\bigotimes_{j=1}^Ne^{-i\theta_j Z/2}$ is canceled.
Therefore, when we calculate the fault-tolerant
threshold of our protocol, we can assume that all $\theta_j=0$ without loss of
generality.
As in Ref.~\cite{Raussendorf_Ann}, we assume
that $|\mathcal{C}_\Theta\rangle$
is created in the stepwise manner (Fig.~\ref{step}).
In our case, however, the additional fifth step is introduced as is shown
in Fig.~\ref{step} (e).
First, let us calculate the single-qubit $Z$ error probability 
$\lambda_{j}$ $(j=1,2,3)$
on each of the three qubits labeled with $1,2,3$ in Fig.~\ref{bcc3} (b)
after creating $|\mathcal{C}_\Theta\rangle$.
By a straightforward calculation,
we obtain 
$\lambda_1=32p_2/15+8p_2/15+2p_P/3$,
$\lambda_2=16p_2/15+2p_P/3$,
and
$\lambda_3=8p_2/15+2p_P/3$
up to the first order of $p_P$ and $p_2$.
Once an erroneous $|\mathcal{C}_\Theta\rangle$ is created,
we start local measurements.
As is shown in Fig.~\ref{tower3}, the three qubits are
sequentially measured in numerical order.
Such a sequential measurement propagates all pre-existing
errors on qubits labeled with 1 and 2 to
the qubit labeled with 3. 
By a straightforward calculation, the accumulated error probability
on the qubit labeled with 3
by such a propagation
is
$\lambda_{total}\equiv\lambda_1+\lambda_3+2\times2p_M/3$ for 
$(\phi_1,\phi_2,\phi_3)=(0,0,0)$.
(We have only to consider the measurement pattern which corresponds
to the effective $X$ measurement on the qubit labeled with 1 in
Fig.~\ref{bcc3} (a),
because we are now interested in the topological error-correction of
the bulk qubits.) 
The value $\lambda_{total}$ 
corresponds to the quantity $q_1$, which was studied in 
Ref.~\cite{Raussendorf_Ann}, 
of the qubit labeled with 1 in Fig.~\ref{bcc3} (a).
The correlated two-qubit error probability~\cite{Raussendorf_Ann}
$q_2=4p_2/15+O(p_2^2)$ 
in our protocol
is the same as that in Ref.~\cite{Raussendorf_Ann}.
If we assume $p_P=p_M=p_2=p$,
we obtain the bulk topological
error threshold as $p=4.3\times10^{-3}$
from Fig.~10 of Ref.~\cite{Raussendorf_Ann}
where the threshold curve of $(q_1,q_2)$
is numerically calculated by using the minimum-weight-perfect-matching
algorithm.

The primal defects are created by measuring 
the edge qubits inside the defect region in the $Z$ basis.
At the boundary of the defect region (surface of the defects),
these measurements introduce additional $Z$ errors 
on the face qubits (i.e., dual qubits), 
which has an effect of decreasing the threshold value.
At the same time, the existence of the defects changes the boundary 
condition of the bulk region;
at the surface of the primal defects, the dual lattice has 
a smooth boundary.
Thus, there is excess syndrome
available at the defect surface, which has an effect of increasing 
the threshold value.
In Ref.~\cite{Raussendorf_Ann},
they have performed numerical simulation for lattices of 
size $L \times L \times 2L$,
where half of the lattice belongs to the bulk region $V$ and the 
other half to the defect region $D$.
In their calculations, the error probability of the dual qubits 
of the surface of the defect
is doubled to investigate the surface effect.
Their numerical result indicates that 
while the surface effect due to the smooth boundary 
(i.e., increasing the threshold value) is noticeable,
the intersection point of fidelity curves is slowly converging 
to the threshold value  
of the bulk region in the increasing number of the lattice size.
This indicates that the $Z$ basis measurements for the defect creations 
do not lower the threshold for TMBQC.
In the present case, the error probability of $Z$ basis measurement
is increased by $\lambda _Z = 2(2p_M/3)+2(2p_P/3)+p_2$ due to 
the additional qubits
for the blind $Z$ basis measurement.
However, when $p_M=p_P=p_2=4.3 \times 10^{-3}$,
the error probability $\lambda _Z =1.6 \times 10^{-2}$
is still smaller than the situation that has been considered 
in Ref.~\cite{Raussendorf_Ann}.
Thus the defect creations 
do not lower the threshold in the blind setup again, 
and hence the threshold value
is determined by that for the bulk region.

Topological error correction breaks down near the singular qubits,
and it results in an effective error on the 
singular qubits~\cite{Raussendorf_PRL,Raussendorf_NJP}.
This effective error is local because singular qubits are well separated
from each other.
Magic state distillation~\cite{magic} can tolerate a rather large amount
of noise, and therefore the overall threshold is determined by
that for the bulk topological region~\cite{Raussendorf_PRL,Raussendorf_NJP}.
In fact,
the recursion relations of the distillations of 
$Y$ and $(X+Y)/\sqrt{2} \equiv T$ basis states 
are given by $\epsilon ^{Y}_{l+1}= 7 
(\epsilon ^{Y}_{l})^3 + \epsilon _{top}^{Y}$ 
and $\epsilon ^{T}_{l+1}= 35 (\epsilon ^{T}_{l})^3+ 
\epsilon _{top}^{T}$,
where $\epsilon _{top}^{Y}$ and $\epsilon _{top}^{T}$ 
indicate the probability
of errors introduced by the Clifford gates for
the magic state distillation~\cite{Raussendorf_NJP}.
In order to optimize their overheads,
the scale factor and defect thickness at each distillation level
are chosen so that 
$\epsilon _l^{Y,T}$ and $\epsilon _{top}^{Y,T}$ are balanced.
Since $\epsilon _{top}^{Y}$ and $\epsilon _{top}^{T}$
can be reduced rapidly by increasing the scale factor 
and the thickness of the defect,
the threshold values for the magic state distillation can be 
determined as $\epsilon ^Y = 0.38$ and $\epsilon ^{T}=0.17$
for $Y$- and $T$-state distillations,
respectively.
In the present decorated case,
the error probability of the injected magic state 
is at most $\epsilon=4p_2 + 2p_2 +3(2p_M/3) +3 (2p_P/3) $,
where the first, second, third, and forth terms indicate four 
C$Z$ gates for generating RHG lattice,
two C$Z$ gates for the decoration, three measurements, 
and three state preparations.
With $p_M=p_P=p_2=4.3 \times 10^{-3}$,
$\epsilon = 10p_2=0.043$ which is sufficiently 
smaller than the threshold values 
for the magic state distillations.

In the above arguments, we have assumed 
$p_P=p_M=p_2$ for simplicity.
However, $p_P$ might be much larger than $p_M$ and $p_2$,
since Alice's quantum technology is assumed to be much weaker
than that of Bob, and the randomly-rotated qubits are sent from
Alice to Bob through a probably noisy quantum channel.
(In addition, Alice cannot distill her qubits, since she cannot
use any two-qubit gate.)
Hence let us consider another representative scenario, 
$p_P=10p$, $p_M=p_2=p$.
Interestingly, the direct calculation shows that
the error threshold is $p=1.6\times10^{-3}$ (i.e., 
still of the order of $10^{-3}$).
This suggests the nice robustness 
on Alice's side
in the topological fault-tolerant protocol.
(Note that this result is reasonable 
because the preparation error behaves as an independent error,
and independent errors are known to be easy to correct.
In fact, if there is no correlated error, TMBQC can tolerates
the independent error up to 2.9\verb|%|.)


\section{Discussion}
In addition to the probabilistic depolarizing noise,
which we have considered,
there are many possibilities of errors.
For example, quantum computation can suffer from the detectable qubit loss,
such as photon loss, atoms or ions escaping from traps, or,
more generally, the leakage of a qubit out of the computational
basis in a multilevel system.
In Ref.~\cite{Barrett_loss}, 
the threshold of the TMBQC for the qubit loss was studied.
Here let us briefly explain that we can obtain a similar threshold for
the qubit loss in our blind protocol.
As in Ref.~\cite{Barrett_loss},
let us assume that losses are independent and identically
distributed events with the probability $p_{loss}$.
If one of the three qubits labeled with 1, 2, and 3 in Fig.~\ref{bcc3} (b)
is lost, then we just consider
the entire three-qubit chain is lost.
Then, we can use the result of Ref.~\cite{Barrett_loss},
and our threshold for the qubit loss is obtained by replacing
their $p_{loss}$ with $3p_{loss}$.
Note that if we also use the post-selected scheme of Ref.~\cite{Barrett_loss},
then the overhead is $\simeq(1-3p_{loss})^{d^3}$,
which is still independent of the size of the algorithm and
hence ensuring scalability.
Another possible error, the non-determinism of 
C$Z$ gates, was considered in Refs.~\cite{Fujii,Li}
for TMBQC.
For example, in Ref.~\cite{Fujii}, the three-dimensional resource state
is created by fusing the ``puffer ball" states.
If the one-dimensional chain of two qubits is 
added to the root qubit of
each puffer ball state, $|\mathcal{C}_\Theta\rangle$
can be created by a similar fusion strategy, 
and the threshold can also be calculated
in a similar way.

Although the simulation of fault-tolerant quantum circuits in the blind 
MBQC on the brickwork state was mentioned in Ref.~\cite{blindcluster},
it is only the existence proof.
Neither a concrete scheme nor an explicit calculation of the threshold
was given.
Furthermore,
on the two-dimensional brickwork state, we should
use the fault-tolerant scheme of the one-dimensional nearest-neighbour
circuit model architecture.
For the circuit model, the threshold of this scheme
is of the order of $10^{-5}$~\cite{Stephens1,Stephens2}.
If we implement this scheme on MBQC,
the threshold should be $\sim 10^{-6}$ due to the extra 
qubits~\cite{Dawson1,Dawson2}.
As is mentioned in Ref.~\cite{blindcluster},
the threshold should be increased if the three-dimensional
brickwork state is considered.
However, the explicit calculation of the threshold
for the scheme of 
Ref.~\cite{Gottesman}
is not known and should be smaller than that of TMBQC.

In the protocol,
Alice performs the decoding operation (error correction)
by using the classical data from Bob~\cite{Raussendorf_PRL,Raussendorf_NJP}.
We have calculated the threshold value for the present blind protocol
by following the result in Ref. \cite{Raussendorf_Ann},
where the minimum-weight-perfect-matching algorithm
is used for the decoding problem.
Although the minimum-weight-perfect-matching
is an efficient algorithm in the sense that it scales polynomially,
it might cost large classical computational resources
when the lattice size is large.
However, more efficient classical
algorithms for the decoding problem have also been 
developed~\cite{Fowler3,Poulin1,Poulin2},
one of which~\cite{Fowler3} achieves the decoding of the lattice of 4
million qubits
within a few seconds by using a today's typical classical computer,
while the resultant threshold $0.9\%$ is higher than that $0.75\%$ in
Ref.~\cite{Raussendorf_Ann}.
In this sense, Alice's classical processing
does not present any problem here.

\section{Methods}
\subsection{Definition of the blindness}
Here we show the blindness of our protocol.
Intuitively, a protocol is blind if Bob, given all the classical and quantum information during the protocol, 
cannot learn anything about Alice's computational angles, input and output~\cite{blindcluster,TVE,Vedran}. 
A formal definition we adopt here is as follows. 
(See also Refs.~\cite{blindcluster,TVE,Vedran}.)

A protocol is blind if: 
(B1) the conditional probability distribution of  
Alice's computational angles, 
given all the classical information Bob can obtain during the protocol, 
and given the measurement results of any POVMs which Bob may perform
on his system at any stage of the protocol,
is equal to the a priori probability distribution of
Alice's computational angles;
and (B2) the final classical output is one-time padded
to Bob.

\subsection{Our protocol satisfies B1}
Let us define 
$\Delta\equiv(\Delta_1,...,\Delta_N)$,
$\Phi\equiv(\Phi_1,...,\Phi_N)$,
$\Theta\equiv(\Theta_1,...,\Theta_N)$,
$R\equiv(R_1,...,R_N)$,
where
$\Delta_j,\Theta_j,\Phi_j\in{\mathcal A}\equiv\{\frac{k\pi}{4}|k=0,1,...,7\}$
and 
$R_j\in\{0,1\}$ are random variables, 
corresponding to the angles sent by Alice to Bob, 
random prerotations, 
Alice's secret computational angles, 
and 
the hidden binary parameters, respectively.
From the construction of the protocol, the relation 
$\Delta_j=\Phi_j+\Theta_j+R_j\pi$ $(\mbox{mod}~2\pi)$
is satisfied.
Let $\{\Pi_j\}_{j=1}^{m}$ be a POVM which Bob may perform on 
his $\{|\theta_j\rangle\}_{j=1}^N$. 
Let $O\in\{1,...,m\}$ be the random variable corresponding to the result
of the POVM. Bob's knowledge about Alice's computational angles is given by the conditional probability distribution
of $\Phi=(\phi_1,...,\phi_N)$ given
$O=j$ and  $\Delta=(\delta_1,...,\delta_N)$: 
$P(\Phi=(\phi_1,...,\phi_N)|O=j,\Delta=(\delta_1,...,\delta_N))$.

From Bayes' theorem, we have
\begin{eqnarray*}
&&P\Big(\Phi=(\phi_1,...,\phi_N)
\Big|O=j,\Delta=(\delta_1,...,\delta_N)
\Big)\\
&=&
\frac{P\Big(O=j\Big|\Phi=(\phi_1,...,\phi_N),\Delta=(\delta_1,...,\delta_N)\Big)
P\Big(\Phi=(\phi_1,...,\phi_N),\Delta=(\delta_1,...,\delta_N)\Big)}
{P\Big(O=j,\Delta=(\delta_1,...,\delta_N)\Big)}\\
&=&
\frac{P\Big(O=j\Big|\Phi=(\phi_1,...,\phi_N),\Delta=(\delta_1,...,\delta_N)\Big)
P\Big(\Phi=(\phi_1,...,\phi_N)\Big)
P\Big(\Delta=(\delta_1,...,\delta_N)
\Big)}
{P\Big(O=j\Big|\Delta=(\delta_1,...,\delta_N)\Big)
P\Big(\Delta=(\delta_1,...,\delta_N)\Big)}\\
&=&
P\Big(\Phi=(\phi_1,...,\phi_N)\Big)
\frac{\mbox{Tr}\Big[\Pi_j\bigotimes_{i=1}^N
\frac{1}{2}\sum_{r_i=0}^1
|\delta_i-\phi_i-r_i\pi\rangle\langle
\delta_i-\phi_i-r_i\pi|
\Big]}
{\mbox{Tr}
\Big[\Pi_j
\bigotimes_{i=1}^N
\frac{1}{8}\frac{1}{2}\sum_{\phi_i\in{\mathcal A}}
\sum_{r_i=0}^1
|\delta_i-\phi_i-r_i\pi\rangle\langle
\delta_i-\phi_i-r_i\pi|
\Big]}\\
&=&P\Big(\Phi=(\phi_1,...,\phi_N)\Big).
\end{eqnarray*}

\subsection{Our protocol satisfies B2}
It is easy to confirm that
when Bob measures the qubit labeled with 3 in Fig. 1 (b),
the state is one-time padded with $Z^{s_1}X^{s_2}$,
where $s_1$ ($s_2$) is the measurement
result of the qubit labeled with 1 (2) in Fig. 1 (b).
The values of $s_1$ and $s_2$ are unknown to Bob, since $\{r_j\}_{j=1}^N$ are
unknown to Bob.
We can show that $\{r_j\}_{j=1}^N$ are unknown to Bob as follows.
\begin{eqnarray*}
&&P\Big(R=(r_1,...,r_N)
\Big|O=j,\Delta=(\delta_1,...,\delta_N)
\Big)\\
&=&
\frac{P\Big(O=j\Big|R=(r_1,...,r_N),\Delta=(\delta_1,...,\delta_N)\Big)
P\Big(R=(r_1,...,r_N),\Delta=(\delta_1,...,\delta_N)\Big)}
{P\Big(O=j,\Delta=(\delta_1,...,\delta_N)\Big)}\\
&=&
\frac{P\Big(O=j\Big|R=(r_1,...,r_N),\Delta=(\delta_1,...,\delta_N)\Big)
P\Big(R=(r_1,...,r_N)\Big)
P\Big(\Delta=(\delta_1,...,\delta_N)
\Big)}
{P\Big(O=j\Big|\Delta=(\delta_1,...,\delta_N)\Big)
P\Big(\Delta=(\delta_1,...,\delta_N)\Big)}\\
&=&
P\Big(R=(r_1,...,r_N)\Big)
\frac{\mbox{Tr}\Big[\Pi_j\bigotimes_{i=1}^N
\frac{1}{8}\sum_{\phi_i\in{\mathcal A}}
|\delta_i-\phi_i-r_i\pi\rangle\langle
\delta_i-\phi_i-r_i\pi|
\Big]}
{\mbox{Tr}
\Big[\Pi_j
\bigotimes_{i=1}^N
\frac{1}{8}\frac{1}{2}\sum_{\phi_i\in{\mathcal A}}
\sum_{r_i=0}^1
|\delta_i-\phi_i-r_i\pi\rangle\langle
\delta_i-\phi_i-r_i\pi|
\Big]}\\
&=&P\Big(R=(r_1,...,r_N)\Big)\\
&=&\frac{1}{2^N}.
\end{eqnarray*}




{\bf Acknowledgement} We acknowledge supports from JSPS, ANR
(StatQuant, JC07 07205763), and MEXT(Grant-in-Aid for Scientific 
Research on Innovative Areas 20104003).

{\bf Author contributions} 
Both authors contributed equally to this work.

{\bf Author information} Correspondence and requests for materials
should be addressed to T.M. (morimae@gmail.com).
The authors declare no competing financial interests.

\end{document}